\documentclass[%
 reprint,onecolumn,
 amsmath,amssymb,
 aps,
]{revtex4-2}

\usepackage{graphicx}
\usepackage{dcolumn}
\usepackage{bm}
\usepackage[utf8]{inputenc}
\usepackage[T1]{fontenc}
\usepackage{mathptmx}
\usepackage{etoolbox}
\usepackage{graphicx}
\usepackage{subcaption}
 \usepackage{amsmath}
 \usepackage{overpic}
\usepackage{caption}
\usepackage{graphicx}
\usepackage{dblfloatfix}
\usepackage{placeins}
\usepackage{ragged2e}

\begin{document}

\preprint{APS/123-QED}

\title{Magneto-coalescence phenomena in colliding ferrofluid droplets}





\author{Neeladri Sekhar Bera$^{1}$}
\email{neeladrisekharb@gmail.com}
\author{Abhishek Kumar Jaiswal$^{1}$}
\author{Devranjan Samanta$^{2}$}
\author{Purbarun Dhar$^{1}$}
\email{purbarun@mech.iitkgp.ac.in}

\affiliation{$^{1}$Hydrodynamics and Thermal Multiphysics Lab (HTML),
Department of Mechanical Engineering,
Indian Institute of Technology Kharagpur,
West Bengal -- 721302, India}

\affiliation{$^{2}$Department of Mechanical Engineering,
Indian Institute of Technology Ropar,
Punjab -- 140001, India}


\begin{abstract}
We explore the collision hydrodynamics of a ferrofluid droplet falling freely onto a sessile droplet of the same liquid, in the presence of a horizontal magnetic field; a configuration that couples droplet-on-droplet coalescence with concomitant field-governed wetting and spreading. Using high-speed imaging, we track the events through crown formation, radial spreading, and rim detachment (under specific conditions), across three ferrofluid concentrations, two substrates of different wettability (glass and PET), and a range of impact velocities and magnetic field strengths. The maximum crown height is noted to scale as $H_{c,\max}/D_t\sim Fr^{0.5}$ ($Fr:$ Froude number); well below the ballistic upper bound of $H_{c,\max}/D_t\sim Fr$. At zero-field, the maximum spreading collapses onto the boundary-layer scaling $\beta_{0,\max}\sim (We_0/Oh)^{1/6}$ ($We_0$: Weber number, $Oh:$ Ohnesorge number) when expressed in terms of the merged impact velocity, and coalesced-drop size . With the field applied, a bulk-dissipation energy balance predicts $\beta_{\max}\sim Z^{1/5}$, where $Z$ combines the magnetic-driving,
and inertial-capillary-viscous terms, but the observations instead follow a markedly weaker $\sim Z^{1/11}$, a deficit traced to enhanced dissipation from the magnetoviscous effects, and manifested via an effective Ohnesorge number. Finally, rim detachment occurs beyond a field- and height-dependent threshold, described by a size-independent criterion $Fr^2Bo\approx 3550$ ($Bo:$ Bond number), above which the rim may fragment into daughter droplets. These scalings provide predictive tools for magnetically assisted printing, droplet-on-demand systems, and coating processes, where repeated droplet collisions occur on the residual liquid droplet or layer.
\end{abstract}

\maketitle

\section{Introduction}
\label{sec:intro}

Every time a droplet lands on a surface, a violent, albeit short-lived, competition plays out between inertia, surface tension, and viscosity. The outcome of the competition decides whether the droplet spreads out neatly, throws up a crown, breaks up into smaller droplets, or bounces away entirely. This few-millisecond long event turns out to matter a very great deal in practice and in utilities; in addition to the innately rich physics and the involved scientific curiosity. Ink-jet printing (extensively used nowadays for printed additive manufacturing, for fabrication of printed and flexible electronics, etc.) needs a drop to spread to a
predictable footprint and be arrested therein~\cite{derby2010inkjet}; spray cooling and quenching depend on how well a stream of drops wets and replenishes a liquid layer on a hot
surface~\cite{breitenbach2018drop}; and spray coating and pesticide deposition need drops to spread and stay put, rather than rebound off the surface they were meant to cover~\cite{bergeron2000controlling}. It is the weight of this practicality, as much as the fluid mechanics itself, that has made drop impact on a solid one of the most intensively studied
problems in the field, with the main regimes of spreading, receding, and splashing now mapped out in detail~\cite{yarin2006drop,josserand2016drop}.
What these classical studies share, however, is a dry, empty surface waiting for the drop to interact. In most of the applications mentioned above, that is not the reality: a spray deposits drops one after another, a print-head fires repeatedly over the same spot, and a coating builds up pass by pass, and thus a falling drop typically lands on the same liquid that a predecessor droplet or film has left behind an instant back.
 
In conjunction to this classical picture, a different question has been gaining attention: can the outcome of such an impact be steered, and not just observed? Ferrofluids make this a possible outcome. They are ordinary Newtonian liquids, with magnetic nanoparticles suspended in them, forming a stable colloidal system. In the presence of a magnetic field, the fluid is magnetized and, with it, experiences a body force that an ordinary liquid does not feel~\cite{rosensweig2013ferrohydrodynamics}. Since this force can be turned on/off, modulated, and aligned to a desired direction, ferro-hydrodynamics of ferrofluid drop impact has become a natural way to test whether a field can actively control wetting/spreading dynamics. On rigid surfaces, a vertical magnetic field has been reported to arrest the rate of spreading, and increase dissipation as the field strength grows~\cite{zhou2019effects}, and on superhydrophobic rigid
surfaces, it can stop a droplet from rebounding altogether, and decides whether the rim breaks apart or stays intact during the impact events~\cite{sahoo2021collisional,sahoo2022vertical}. In arc welding a magnetic field is used to guide the molten metal drop and cut down on the spatter they leave behind~\cite{chang2014impacts}. Even within reports that study a vertical magnetic field, the details matter: adding a vertical gradient on top of the magnetic field introduces a Kelvin body force that pulls the drop downward as it spreads, and this has been shown to reduce the maximum spreading diameter through two competing magnetic
effects; one tied to the field's ability to do magnetic work on the fluid, and the other to how it modifies the effective surface energy of the droplet~\cite{li2022maximum}. A horizontal magnetic field has been used to suppress the rebound of non-Newtonian ferrofluid droplet upon impacting on superhydrophobic surfaces~\cite{vvs2021magneto}. Research on liquid-metal drop impact, where a horizontal magnetic field disrupts the circular symmetry of spreading, is shown to stretch the droplet lamella into an ellipse along the field direction (however, here, the elongation is driven by an induced Lorentz force, unlike the magnetization body force at work in a ferrofluid)~\cite{yang2018elliptical}. Further, energy-balance models, theorized for both diamagnetic and paramagnetic
ferrofluids, indicate that the field reconfigures the manner in which the kinetic energy of the colliding droplet is distributed between spreading, dissipation, and surface deformation components ~\cite{ahmed2018maximum,ahmed2018effects}. Hence, the field does not act only as a small correction force to drop impact events; rather it is a second major control knob, in addition to the impact inertia, and the fluid properties.
 
This second knob is easiest to realize once we know which system it is being incorporated to. For a droplet falling on a dry surface, the maximum spreading diameter is established by a balance of energy, and the two limiting cases are well understood. For low-viscosity fluids, surface tension eventually arrests the spreading process. The simplest argument treats all of the kinetic energy of the droplet at impact as fully transformed into surface energy at the moment of maximum spreading, and this balance yields $\beta_{max} \equiv D_{max}/D_i \sim \mathrm{We}^{1/2}$, where the Weber number $\mathrm{We} = \rho V_i^2 D_i/\sigma$ compares inertia to surface tension~\cite{collings1990splat,bechtel1981impact}. Here $\rho, \sigma, V_i, D_i$ are the fluid density, surface tension, impact velocity, and initial diameter of the impacting drop, respectively. Clanet \textit{et al.}~\cite{clanet2004maximal} later showed that a
more careful argument, based on how the drop decelerates as it spreads, rather than on energy conservation alone, yields  $\beta_{max} \sim \mathrm{We}^{1/4}$ instead, and that this expression better agrees with experiments. If viscous dissipation within the spreading film arrests the spreading, then energy balance yields $\beta_{max} \sim \mathrm{Re}^{1/5}$, where the Reynolds number $\mathrm{Re} = \rho V_i D_i/\mu$ compares inertia to viscosity~\cite{chandra1991collision,pasandideh1996capillary,scheller1995newtonian,roisman2002normal}, where $\mu$ is the dynamic viscosity of the fluid. 

Most real impact scenarios exist somewhere between these two extrema, and Laan
\textit{et al.}~\cite{laan2014maximum} showed that a single
expression, interpolating between the $\mathrm{Re}^{1/5}$ scaled viscous limit, and the $\mathrm{We}^{1/2}$ scaled capillary limit, reproduces the spreading of many different liquids across this whole range. A more recent refinement sharpens this scenario further by tracing the viscous dissipation specifically to a thin boundary layer
near the wall~\cite{liu2025maximum}. Once a magnetic field is incorporated, this tidy, two-regime mechanics is no longer sacrosanct. Depending on the philosophy of the study, the field has been introduced in as an extra energy term sitting alongside the Weber and Reynolds numbers~\cite{li2022maximum}, or treated as a parameter that morphs the fluid's effective viscosity and
dissipation term directly~\cite{ahmed2018maximum,zhou2019effects}. It is noteworthy that these explorations rest almost entirely on vertical magnetic field scenarios, and the behavior induced by a horizontal magnetic field has barely been examined for ferrofluid droplet dynamics at all. Which of these is the right way to think about the problem, and whether the same crossover idea that works without a field survives once one is switched on; remains an open question.
 
All of this, though, describes a drop hitting a dry, rigid surface. Land the drop on a puddle or, more precisely, on another sessile drop resting on that surface, and the physics changes in ways that a dry-surface energy balance formulation cannot capture. Damak and Varanasi~\cite{damak2018expansion} showed that when a falling drop merges with a sessile drop of the same liquid, the retraction that follows is not simply the retraction of a bigger single drop. Rather, a viscous regime appears and the retraction dynamics has no counterpart in ordinary single-drop impact, because the two drops must first coalesce and share momentum before anything else may occur. Jaiswal
and Khandekar~\cite{jaiswal2021drop} explored this further on a
superhydrophobic surface, showing that whether the pair coalesces, bounces, or breaks apart, depends sensitively on their relative size and how far off-center the impact is. Pandey and Kondaraju~\cite{pandey2025experimental} recently confirmed, across a range of viscosities, that even the maximum spreading diameter of the
merged pair cannot simply be inferred by plugging the combined volume into a single-drop formula. Put these results together and drop-on-drop impact stops looking like a minor variation on drop-on-solid impact; it is infact a distinct problem, with its own momentum transfer and its own energy balance formalism. That leaves an obvious question unanswered: nearly every spray, print, or coating process deposits
drops onto liquid left behind by earlier drops, and a magnetic field has already been shown to be a strong, controllable handle on how a single ferrofluid drop spreads. So what happens when a falling ferrofluid drop lands on a sessile ferrofluid drop, while within the presence of a magnetic field? To the best of our knowledge, this question remains unexplored.
 
In this article, we aim to answer that question for a ferrofluid drop falling onto a sessile drop of the same ferrofluid, seated on a solid substrate, within a horizontal magnetic field ambience. We follow the impact events in the order it
actually unfolds. Right after the two drops merge, the liquid is thrown outward and upward as a crown, and we show how the height of that crown departs from the crown-formation picture prevalent for impact on a wetted wall~\cite{roisman2002impact,ma2022multi} once a field is present. As the crown falls back and collapses onto the substrate, the merged drop reaches its maximum spreading diameter. We propose a semi-empirical scaling law for it that extends the viscous-capillary crossover idea to this two-drop geometry, and discusses directly whether the magnetic field is better treated as an extra energy term, or as a change to the fluid's effective dissipation. Finally, we identify the condition (expressed through the governing Froude and Bond numbers) for which the rim of the crown detaches from the collapsing sheet at the moment of maximum crown height, rather than retracting along with it. Crown formation, maximum spreading, and rim detachment are not independent outcomes here; they are three stages of the same event. Treating them together is what enables us to connect the drop-impact scaling framework, built for a single drop on a rigid wall, to the coalescence-driven, field-tunable case that spray and printing processes actually encounter.
 
\section{Experimental Materials and methodology}
\label{sec:method}
Three  deionized water-glycerol mixture -based ferrofluids (FF), loaded with Fe$_2$O$_3$
nanoparticles (30-40 nm diameter, Alfa Aesar, India) were used to span a range of viscosities, while keeping the density and surface tension within a comparable band. The nanoparticles were coated with citric acid (as supplied by the manufacturer), which enhances the colloidal stability. The FF were first mechanically stirred for 30 min and then ultra-sonicated for 1 h by probe-type sonicator (Oscar Ultrasonics, India). Zeta potential (Malvern Instruments, USA) values lay outside the range of $-30 mV$ to $ +30 mV$, indicating sufficient colloidal stability. The FFs were found to be stable over timescales significantly longer than those of the experiments. The equilibrium surface tension of the FF was measured using the pendant-drop method and image analysis (using ImageJ software). In Table~\ref{table:property} the composition of different FF are given as $M$ Gly $N$ Fe$_2$O$_3$, where $M$ denotes the v/v\% of glycerol and $N$ denotes the .wt\% of Fe$_2$O$_3$ nanoparticles. The initial magnetic susceptibility $\chi$ of the ferrofluids was taken to correspond to the saturation-magnetization values reported for
Fe$_2$O$_3$-nanoparticle ferrofluids of comparable loading by Rigoni
\textit{et al.}~\cite{rigoni2016static}, following the same convention
used in our companion study of magneto-extensional filament
dynamics~\cite{bera2026non}. To isolate the role of wettability, two solid substrates of contrasting hydrophilicity were used: glass, with a static contact
angle of $15$--$18^{\circ}$, and polyethylene terephthalate (PET), with a static contact angle of
$55$--$60^{\circ}$. Both substrates were mounted
horizontally on a leveled stage and replaced/cleaned between experimental runs to avoid contamination from residual ferrofluid.

\begin{figure*}[]
    \centering
 \begin{minipage}[t]{0.485\textwidth}
        \centering
        \includegraphics[width=\textwidth]{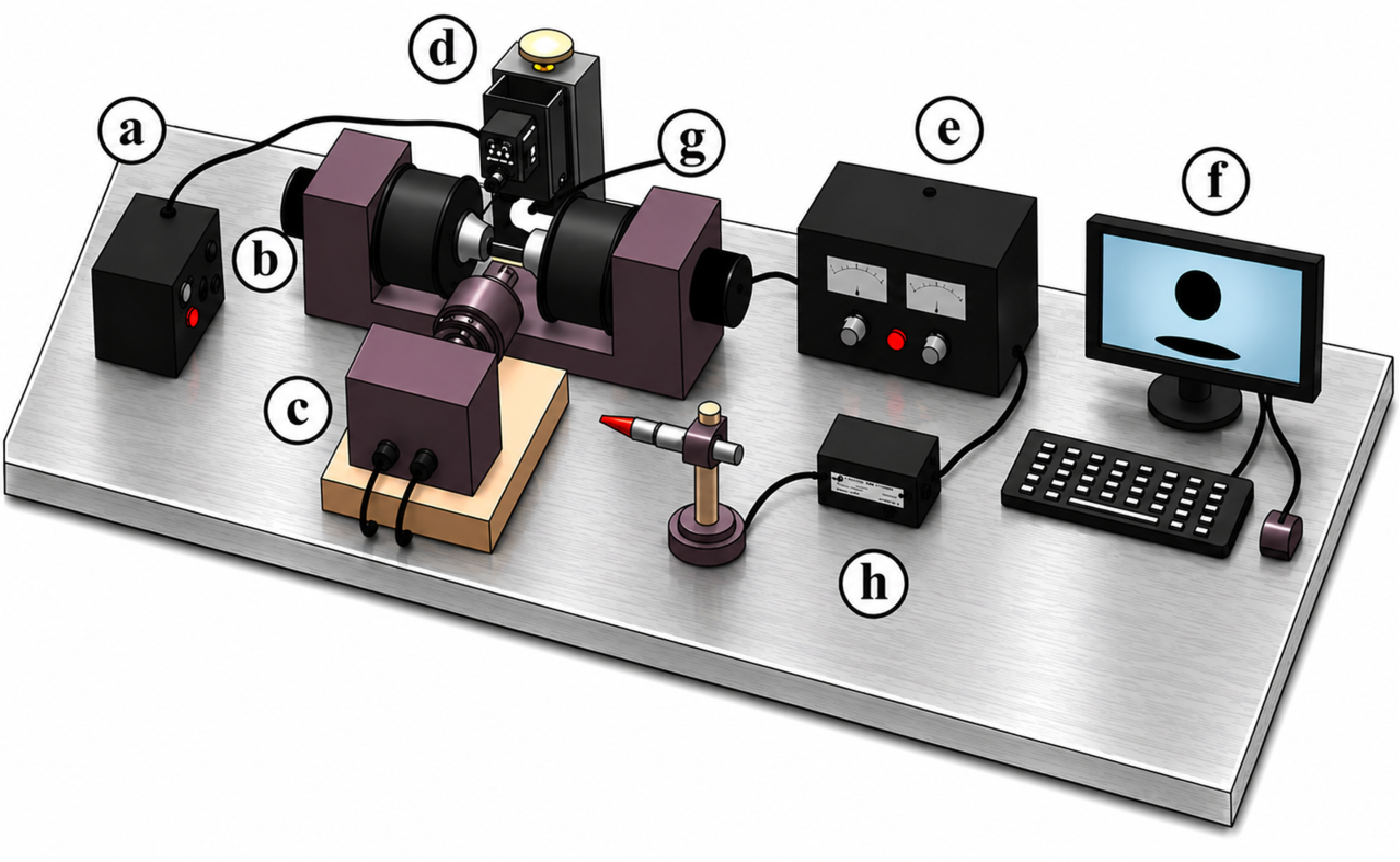}

        \vspace{0.2em}

        \textbf{(a)}
    \end{minipage}
    \hfill
    \begin{minipage}[t]{0.485\textwidth}
        \centering
        \includegraphics[width=1.1\textwidth]{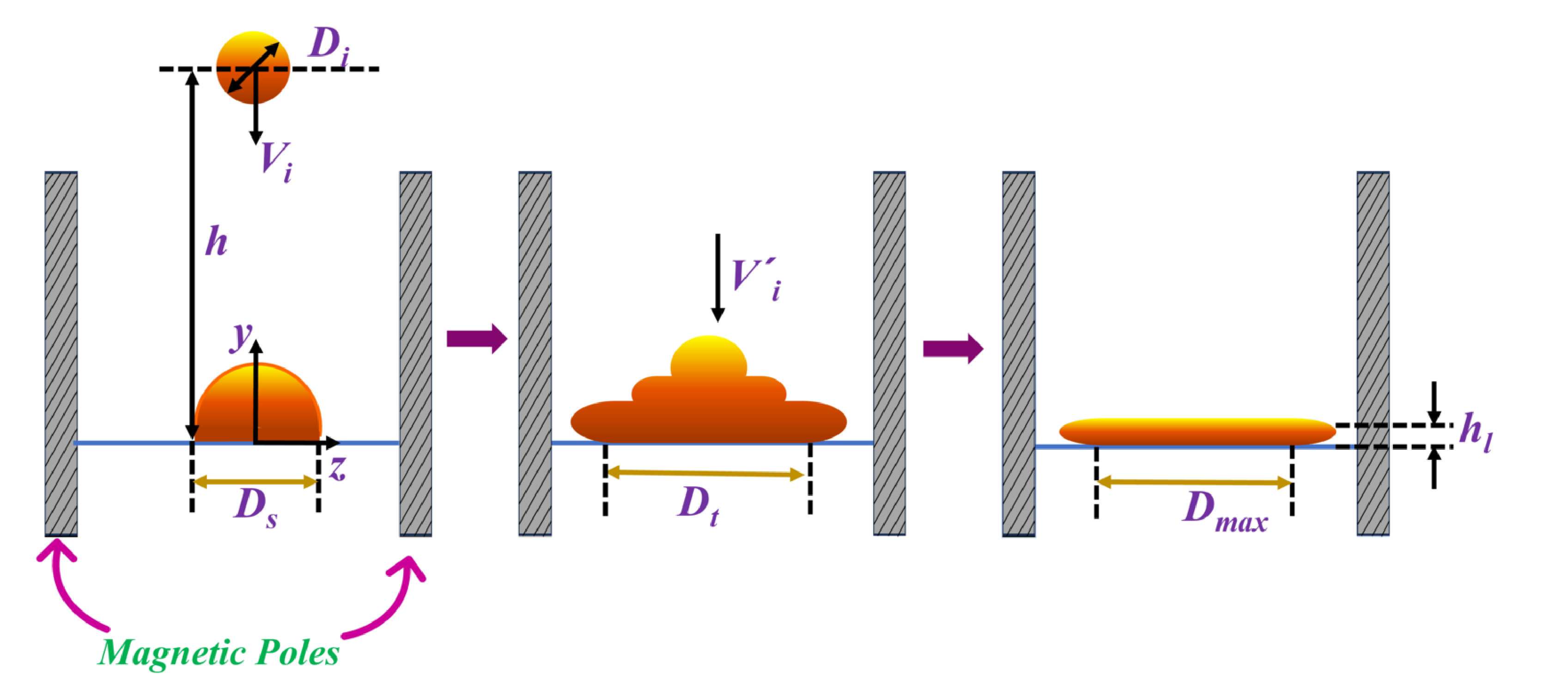}

        \vspace{0.2em}

        \textbf{(b)}
    \end{minipage}

    \vspace{0.5em}

    \begin{minipage}[t]{0.98\textwidth}
        \centering
        \includegraphics[width=\textwidth]{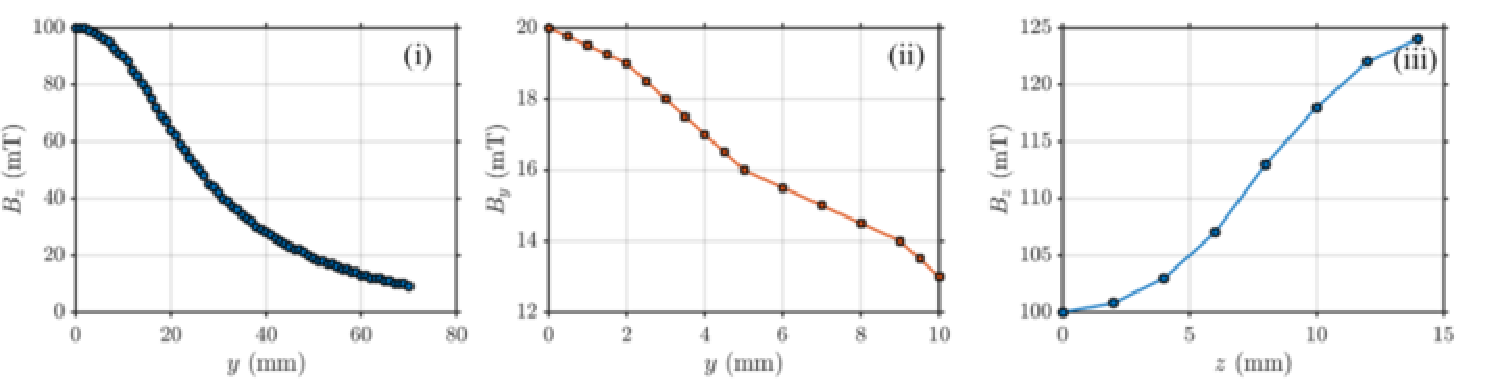}

        \vspace{0.2em}

        \textbf{(c)}
    \end{minipage}

    \caption{\protect\justifying
\textbf{(a)} Schematic of the experimental setup: (a) droplet dispensing system controller, 
(b) electromagnet, (c) high-speed camera, (d) digitized micro-droplet dispenser, 
(e) programmable power supply for the electromagnet, (f) computer system for camera 
control and data acquisition, (g) continuous light source, and (h) gaussmeter. 
\textbf{(b)} Schematic illustration of the drop coalescence process, showing the impacting 
drop of diameter $D_i$, impact velocity $V_i$, initial sessile-drop diameter $D_s$, 
and the subsequent evolution of the coalesced drop with characteristic diameter 
$D_t$ and maximum spreading diameter $D_{\max}$. 
\textbf{(c)} Measured variation of the axial magnetic-field component $B_z$ and radial 
magnetic-field component $B_y$ along the $y$-direction at $z=0$. The magnetic 
fields were measured with $B_z=100~\mathrm{mT}$ at the center.
}

    \label{fig:experimental_setup_magnetic_field}
\end{figure*}

 As shown in Fig.~\ref{fig:experimental_setup_magnetic_field}a, the magnetic field was generated between the poles of an electromagnet (Holmarc, India), powered by a programmable power supply (Polytronic Corp., India). The field strength was measured and calibrated at the sample location using a GaAs sensor-based Hall-effect gaussmeter
(Holmarc, India) by varying the input current. Unlike the vertical-pole, coaxial configuration used in most prior ferrofluid drop-impact studies~\cite{li2022maximum,zhou2019effects}, the
electromagnet poles here were oriented horizontally, with the substrate placed in the gap between them so that the field acts parallel to the substrate and perpendicular to the direction of drop fall. A FF drop was first dispensed gently onto the substrate, at the centre of the pole gap (maintained at $30 mm$
throughout), using a digitized micro-droplet dispenser system (Holmarc, India) fitted with a 24G blunt-tipped needle (DispoVan), to form the sessile target drop; this drop was allowed to settle to rest before the magnetic field was switched on to the desired strength. With the field established, a second ferrofluid drop of the same composition was dispensed from an identical needle positioned
directly above the sessile drop and released to fall, impacting the sessile drop head-on (see Fig.~\ref{fig:experimental_setup_magnetic_field}b). The release height $h$, measured
from the needle tip to the sessile drop, was varied between $9.6$ and $32$~cm to control the impact velocity. This range was set by two practical constraints at the extremes of the parameter space: above $h \approx 32$~cm, the falling drop could no longer be reliably guided into a head-on collision with the sessile drop and impacts became increasingly offset, and below $h \approx 9.6$~cm the falling drop was coming under the effect of magnetic field at the releasing time. Above the corresponding upper limit of field strength, the sessile drop itself was pulled towards the pole, before the falling drop arrived. The reported
ranges of $h$ ($9.6$--$32$~cm) and $B$ ($0$--$0.3$~T) therefore represent the accessible experimental window within which a reproducible, head-on drop-on-sessile-drop impact could be obtained.

The impact event was captured using a high-speed camera (Photron, UK) equipped with a 105~mm macro lens (Nikon), backlit by a continuous LED light source. Images were recorded at 4000 frames per second with a resolution of $1280 \times 1024$ pixels. An ImageJ based custom  image-processing routine was used to extract,
from each frame sequence, the initial diameter $D_i$ and impact velocity $V_i$ of the falling drop, the instantaneous crown height and its maximum value $H_{c,max}$, the maximum spreading diameter $D_{max}$ of the coalesced drop pair, and the instant and configuration at which rim detachment from the collapsing crown occurred. All experiments were conducted at a controlled temperature of 24$^{\circ}$C $\pm$ 1$^{\circ}$C. The variation in properties and measurements of the generated ferrofluid droplets is noted within $\pm5\%$. Each set of experiments was repeated three times. 
\begin{table}[h]

\centering
\caption{Physical properties of the Fe$_2$O$_3$ ferrofluids (FF) studied.}
\label{tab:ferrofluid_properties}
\renewcommand{\arraystretch}{1.4}
\setlength{\tabcolsep}{10pt}
\begin{tabular}{cccc}
\hline
\textbf{Composition} & \textbf{$\rho_{FF}$ (kg/m$^3$)} & \textbf{$\sigma_{FF}$ (mN/m)} & \textbf{$\mu_{FF}$ (mPa$\cdot$s)} \\
\hline
45 Gly 5 Fe$_2$O$_3$ (FF1) & 1144 & $67 \pm 2$ & 5.45 \\
67 Gly 5 Fe$_2$O$_3$ (FF2) & 1209 & $65 \pm 2$ & 19 \\
60 Gly 7.5 Fe$_2$O$_3$ (FF3) & 1200 & $66 \pm 2$ & 11.3 \\
\hline
\end{tabular}
\label{table:property}
\end{table}

\section{Results and discussion}

\subsection{Effect of magnetic field on impact velocity $V_i$}
\label{subsec:effect of magnetic field on impact velocity}

\begin{figure}[]
    \centering
    \includegraphics[width=0.6\columnwidth]{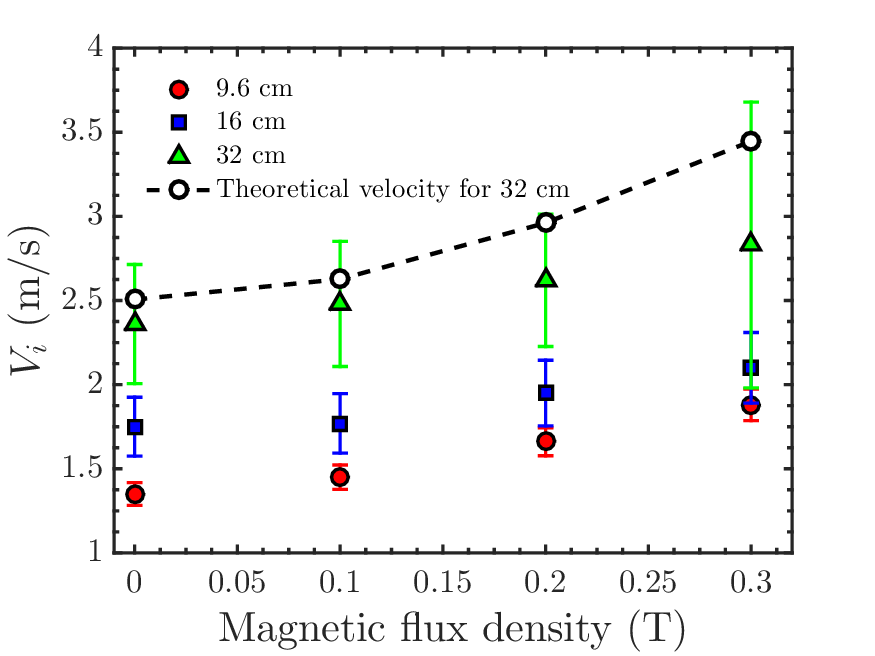}
    \caption{Variation of impact velocity $V_i$ of the impacting FF1 drop just before coalescence with increasing magnetic flux density.}
    \label{fig:Variationofimpactvelwithfield}
\end{figure}

Before we can talk about spreading, crown formation, or rim
detachment, we first need to know how fast the drop is actually moving when it lands under a magnetic field, that is not simply $\sqrt{2gh}$ anymore. When a FF drop falls under the combined
effect of gravity and the magnetic field, its equation of motion can
be written as,
\begin{equation}
\frac{\pi D_i^3}{6}\,\ddot{y}
=
-\frac{\pi D_i^3}{6}\,g
+
\frac{\chi\pi D_i^3}{6\mu_0}\,B\frac{\partial B}{\partial y},
\label{eq:impactvelocity}
\end{equation}
where $D_i$ is the diameter of the impacting drop, $g$ is the
acceleration due to gravity, $B$ is the magnetic flux density, and $\mu_0$ is the magnetic permeability of free space, with the coordinates chosen as shown in Fig.~\ref{fig:experimental_setup_magnetic_field}b. Integrating Eq.~\ref{eq:impactvelocity} from the release point to the instant of
impact gives,
\begin{equation}
\int_{-V_i}^{0} \dot{y}\,d\dot{y}
=
-\int_{h}^{0} g\,dy
+
\frac{\chi}{2\mu_0\rho}
\int_{0}^{h}
\frac{\partial (B^2)}{\partial y}\,dy,
\end{equation}
so that the impact velocity $V_i$ just before coalescence, works out to,
\begin{equation}
V_i^2=2gh+\frac{\chi B^2|_{y=z=0}}{\mu_0 \rho}.
\label{eq:impactveloexpre}
\end{equation}
The second term in Eq.~\ref{eq:impactveloexpre} is where the field enters, and it is worth being precise about where it comes from. Fig.~\ref{fig:experimental_setup_magnetic_field}c shows that the field near the sessile drop is not purely horizontal. Although the dominant component is $B_z$, still a weaker radial component $B_y$ is also present, so a genuine field gradient exists along the direction the
drop is falling, even though $B_y$ itself stays small throughout. It is precisely this gradient that accelerates the falling drop as it approaches the sessile drop, which is why the impact velocity increases with the field strength. $B$ is essentially zero at the
release height $y=h$ and rises to $\sqrt{B_y^2+B_z^2}$ just before impact. Fig.~\ref{fig:Variationofimpactvelwithfield} compares this prediction against the experimentally measured impact velocities  for the FF1 drop,
and the two agree well at lower field strengths. The deviation appears at higher fields, since
Eq.~\ref{eq:impactvelocity} does not account for the aerodynamic drag, the drop experiences over its fall. The impact velocities obtained this way are used throughout the rest of the analysis.

\subsection{Evolution of crown}
\label{subsec:evolution of crown}

The first thing to happen after the falling droplet touches the sessile droplet is not spreading, it is the sudden rise of a thin liquid sheet at the point of contact, which we refer to as the crown. Because
coalescence happens essentially instantaneously compared with the timescales of spreading or crown growth, it can be treated as a momentum-conserving collision. If the falling droplet carries velocity
$V_i$ just before contact, conservation of linear momentum between it and the initially stationary sessile droplet gives,
\begin{equation}
\label{eq:linear momentum}
    m_i V_i = (m_i+m_s)V'_i,
\end{equation}
where $m_i$ and $m_s$ are the masses of the falling and sessile
droplets, respectively. In our experiments the two droplets are of equal
mass, so Eq.~\ref{eq:linear momentum} reduces to $V'_i = V_i/2$. The merged droplet begins its post-coalescence life at exactly half the impact velocity. From this point on, it is $V'_i$, together with a merged length scale $D_t \equiv (D_i^3+D_s^3)^{1/3}$ (with $D_i$ and
$D_s$ the diameters of the impacting and sessile droplets, as defined in
Fig.~\ref{fig:experimental_setup_magnetic_field}b), that set the relevant velocity and length scales for everything that follows. It is worth mentioning that, in our experiments, the impacting droplet diameter was fixed at $D_i\approx2.5mm$ while the sessile droplet $D_s$ varied with the substrate it rested on. Multiplying Eq.~\ref{eq:impactveloexpre} through by $\rho D_t/4\sigma$ turns the impact-velocity relation into a direct statement about the competing forces at play as, $We_m = 1/4(We_0 + Bo_m)$, where $We_m = \rho V_i'^2 D_t/\sigma$ is the magnetic Weber number and $We_0$ is the same inertia-to-surface-tension ratio in the absence of a field. $Bo_m =
\chi B^2 D_t/\mu_0\sigma$ is the magnetic Bond number, comparing the magnetic body force to surface tension. Alongside these, the analysis that follows will also need the Ohnesorge number $Oh = \mu/\sqrt{\rho\sigma D_t}$, which weighs viscous forces against this same inertio-capillary balance. The typical ranges of the above dimensionless numbers in our current experimental setup are $35\le We_0\le 197$, $43\le Bo_m \le 592$ and $0.007\le Oh \le 0.09$ respectively.
 
\begin{figure*}[]
    \centering

    \begin{minipage}[t]{0.75\textwidth}
        \centering
        \includegraphics[width=\textwidth]{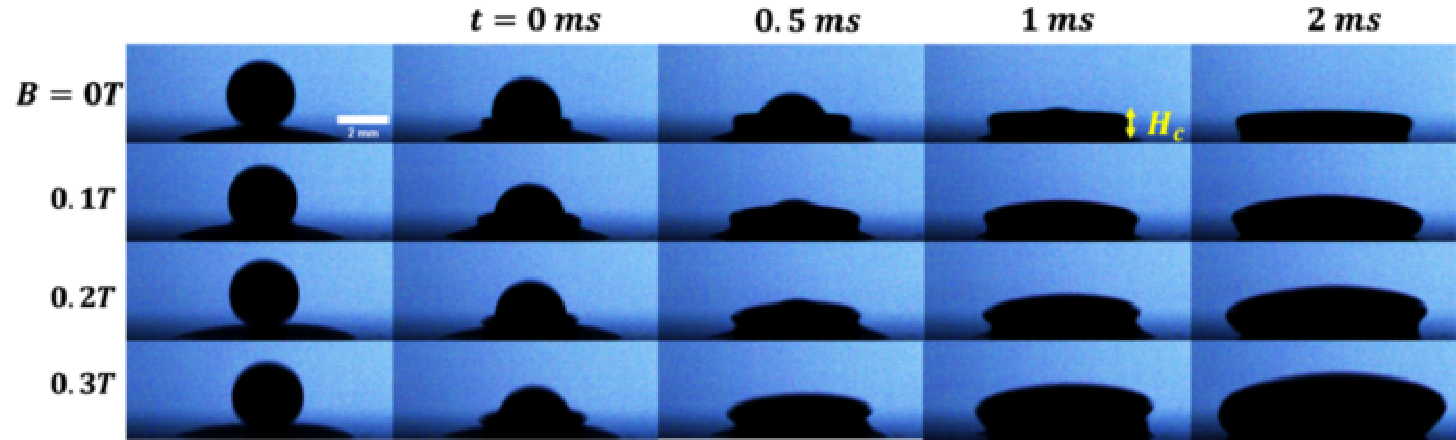}

        \vspace{0.15em}
        \textbf{(a)}
    \end{minipage}
    \begin{minipage}[t]{0.6\textwidth}
        \centering
        \includegraphics[width=\textwidth]{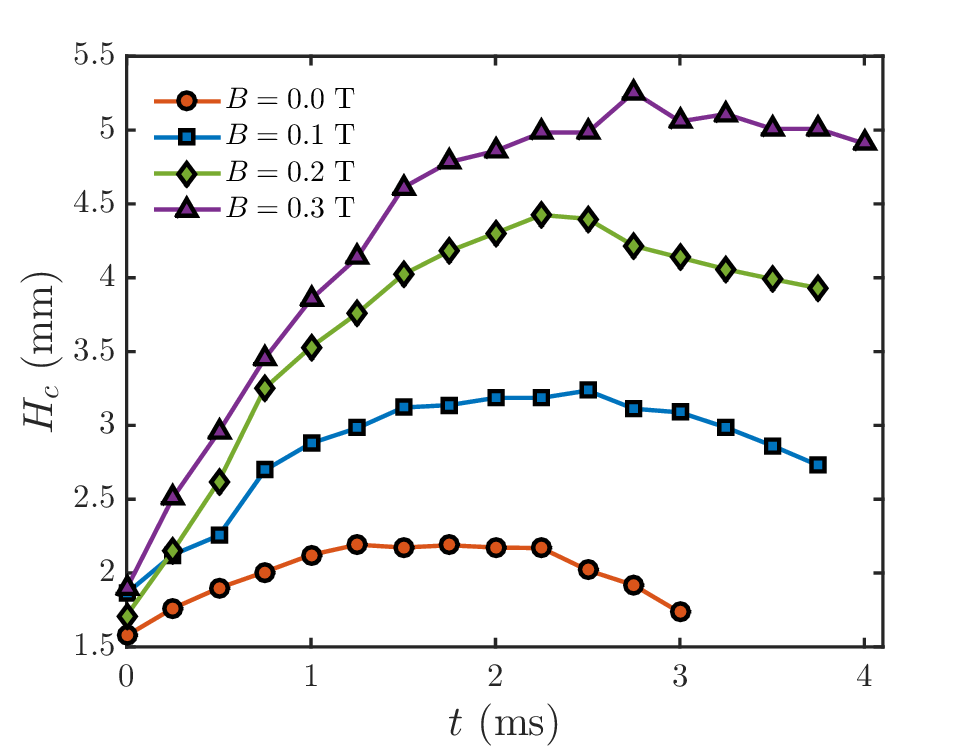}

        \vspace{0.15em}
        \textbf{(b)}
    \end{minipage}

    \caption{\protect\justifying
        (a) Snapshots of evolution of crown height with increasing magnetic field strength. This is the case for the FF1 drop falling from $9.6 cm$ height on the glass substrate. 
    (b) Plot of crown height $H_c$ vs. time $t$ for the same.
}
 \label{fig:crown_height}
\end{figure*}
Fig.~\ref{fig:crown_height} shows the  rise of crown height directly. For a fixed release height, the crown grows taller as the applied field strengthens. This is consistent with what we already established in
Section~\ref{subsec:effect of magnetic field on impact velocity}. A stronger field increases the impact velocity $V_i$, and with it the kinetic energy available to the merged drop, so a larger share of that energy ends up launching the crown.

\subsubsection{Scaling of maximum crown height}
\begin{figure}
    \centering
    \includegraphics[width=0.7\linewidth]{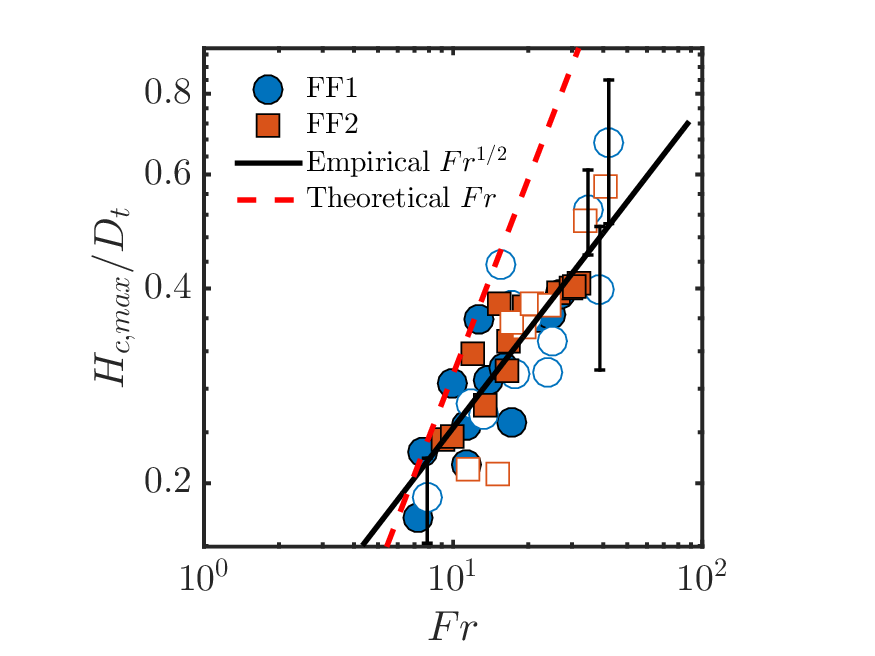}
    \caption{\protect\justifying Scaling for maximum crown height across all heights and magnetic fields. The filled symbols are for the glass substrate, and open symbols are for PET substrate. The equation for the empirical line is $0.08 Fr^{1/2}$ with a $R^2$ value of 0.82. FF3 data also agrees to this empirical scaling, however is not presented here to avoid clustering of too many data points in the plot. }
    \label{fig:crown_height_scaling}
\end{figure}
Before comparing against the measured crown height, it is useful to first establish how high a purely inertial launch could carry it, with nothing else standing in the way. Treating the crown as a
ballistic projectile, launched once, at coalescence, with no further energy input, no surface tension, no viscous loss, and no drag, equate the kinetic energy imparted at launch directly to the potential energy gained at the peak. Taking the crown's characteristic launch velocity $U_c$ to scale with the merged impact
velocity $V_i'$, this balance gives,
\begin{equation}
V'^2_i \sim g H_{c,max}.
\end{equation}
Dividing both sides by $D_t$ turns this into a dimensionless
upper-bound scaling for the maximum crown height,
\begin{equation}
    H_{c,max}/D_t \sim Fr,
\end{equation}
where $Fr$ is the Froude number, defined as $V'^2_i/gD_t$~\cite{wang2023analysis}. This is deliberately the most optimistic version of the story: surface energy, viscous dissipation, and the fact that not all of the impacting mass is actually launched into the crown have all been left out. Accounting for the surface energy properly would require resolving the surface area and thickness of the crown, neither of
which could be reliably obtained with the present experimental setup, and a further complication is that mass visibly accumulates at the rim as the crown grows rather than remaining uniformly distributed. Given these omissions, the real crown height is expected to fall short of
this bound rather than saturate it, and this is indeed what is observed: fitting the measured data across all release heights, magnetic fields, fluids, and substrates, as shown in Fig.~\ref{fig:crown_height_scaling}, yields a universal empirical scaling $H_{c,max}/D_t \sim Fr^{0.5\pm0.02}$, sitting well below the $Fr$ ballistic-derived scaling across the entire range tested. Therefore though the theroretical argument here does not predict the maximum crown height well, yet it sets up an upper bound to it, and the gap between two is a direct measure of how much energy the real crown loses to the mechanism the ballistic picture leaves out. 

\subsection{Effect of magnetic field on maximum spreading}
\label{subsec:effect of magnetic field on maximum spreading}

\begin{figure*}[]
    \centering

    \begin{minipage}[c]{0.47\textwidth}
        \centering
        \includegraphics[width=\textwidth]{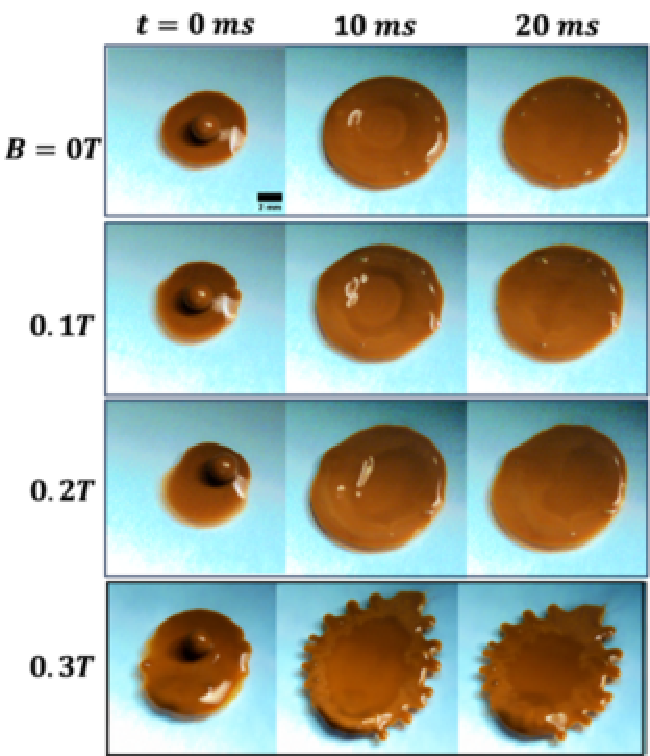}
        
        \vspace{0.1cm}
        \textbf{(a)}
    \end{minipage}
    \hfill
    \begin{minipage}[c]{0.49\textwidth}
        
        \centering
        \includegraphics[width=1.2\textwidth]{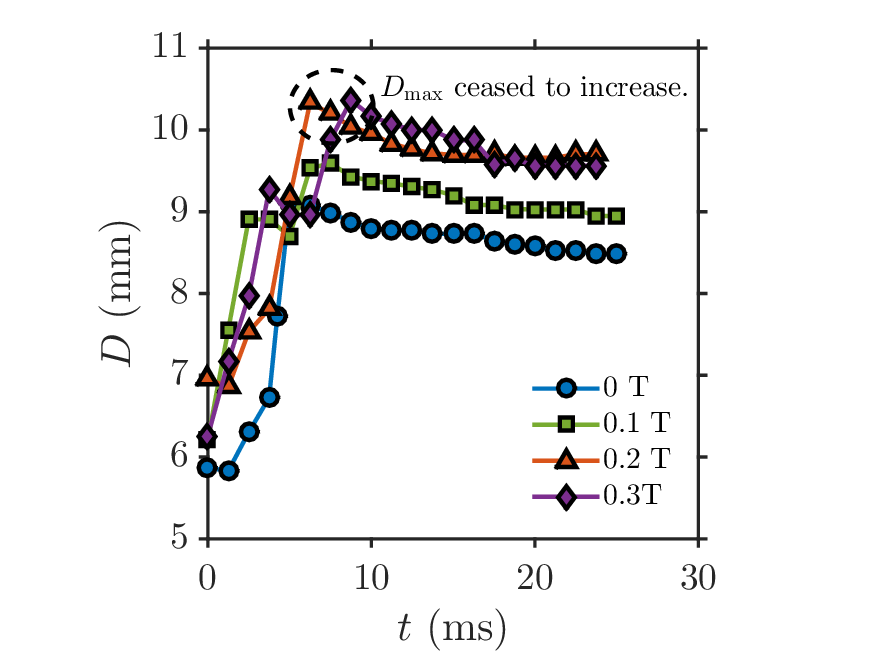}
        
        \vspace{0.05cm}
        \textbf{(b)}
        
        \vspace{0.35cm}
        
        \centering
        \includegraphics[width=1.2\textwidth]{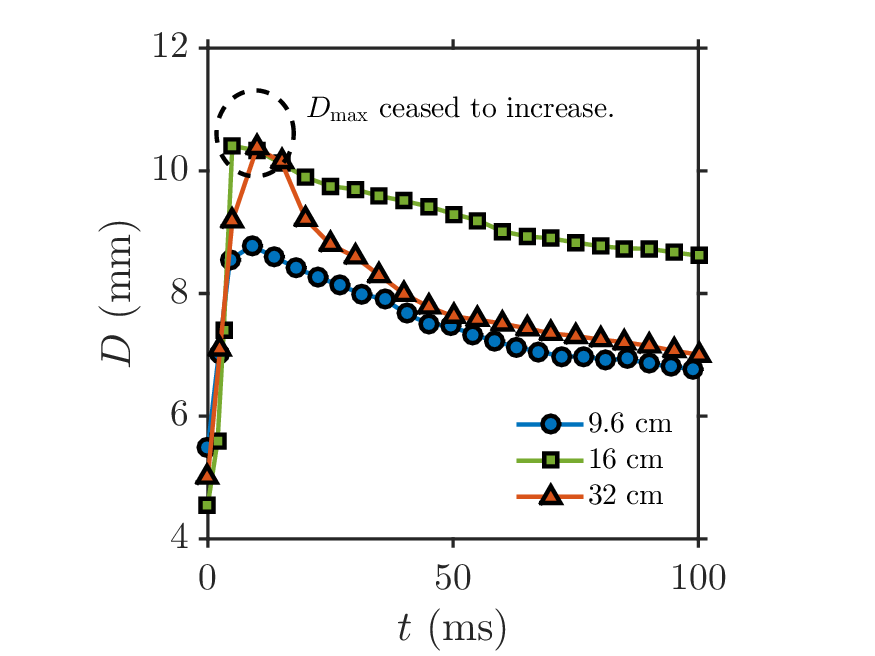}
        
        \vspace{0.05cm}
        \textbf{(c)}
        
    \end{minipage}

    \caption{\protect\justifying
    (a) Snapshots of the evolution of the spread diameter of the FF1 drop falling from a height of $16$ cm onto a glass substrate with increasing magnetic field strength.
    (b) Variation of the spread diameter, $D$, with time for the event shown in (a).
    (c) Variation of the spread diameter, $D$, with time for the FF1 drop falling onto a PET substrate at a magnetic field strength of $0.1$ T for different drop heights.
    }
    \label{fig:spreading_dynamics}
\end{figure*}

Once the crown has collapsed back onto the substrate, the coalesced drop continues to spread outward, and this is where the field's
influence becomes most visible. Fig.~\ref{fig:spreading_dynamics}b
shows the temporal evolution of spreading dynamics at a fixed release height of $16$ cm for increasing field strength. Given the strong field gradient along $z$
that we already show in
Fig.~\ref{fig:experimental_setup_magnetic_field}c, the natural expectation is that the maximum spreading diameter $D_{max}$ should
keep increasing with the field, and up to a point it does. What is more interesting is what happens beyond that point. At higher field
strengths, $D_{max}$ stops growing further and effectively plateaus. The same pattern shows up when the field is held fixed and the release height
is varied instead. Fig.~\ref{fig:spreading_dynamics}c shows that
increasing the height from $9.6$ to $16$ cm raises $D_{max}$, as expected. A higher releasing height signifies greater impact inertia, and more inertia should mean more spreading. But pushing the height
further, to $32$ cm, does not extend this trend; $D_{max}$ again saturates rather than continuing to rise. Both saturations are addressed in the following Sections \ref{subsubsec:Scaling of maximum spreading diameter with magnetic field}, \ref{subsubsec:Rim detachment phenomena and it's criterion} respectively.






\subsubsection{Scaling of maximum spreading diameter at zero magnetic field}
\label{subsubsec:Scaling of maximum spreading diameter at zero magnetic field}

Before the effect of the magnetic field on maximum spreading can be addressed on any quantitative footing, it is necessary to establish how the merged drop behaves in its absence. For a single drop impacting a dry solid, two limiting scalings for the maximum spreading factor are well established: in the capillary limit for low-viscosity fluids, a deceleration-based argument gives $\beta_{max} \sim
We^{1/4}$~\cite{clanet2004maximal}; in the high-viscosity limit, where dissipation is distributed through the bulk of the spreading
lamella, $\beta_{max} \sim Re^{1/5}$~\cite{clanet2004maximal}. Most
Newtonian liquids of practical interest, however, occupy the intermediate regime between these two asymptotes, and it is this regime that governs the present ferrofluids. Scheller and
Bousfield~\cite{scheller1995newtonian} addressed this directly with an extensive experimental study spanning viscosities from 3 to 31
mPa$\cdot$s, and found that the maximum spread radius for a wide range of Newtonian liquids collapses onto a single empirical correlation when plotted against the composite group $We/Oh$,
\begin{equation}
    \beta_{max} \sim  (We/Oh)^{0.166}.
\end{equation}
 Liu \textit{et al.}~\cite{liu2025maximum} recently studied the viscous dissipation in the low-to-moderate viscosity regime as confined to an oscillatory boundary layer of Stokes thickness $\delta \sim D_i Oh^{1/2}$, rather than distributed through the bulk. They derived
\begin{equation}
    \beta_{max} \sim (We/Oh)^{1/6}
\end{equation}
directly from energy conservation. An exponent of $1/6 = 0.166$,
matching the empirical fit of Scheller and Bousfield to within experimental scatter. This boundary-layer scaling was shown to hold
for $Oh \lesssim 0.1$. Together, these results give the clearest picture
currently available of how a single Newtonian drop spreads on a dry solid across the full viscosity range relevant here.
 
\begin{figure}
    \centering
    \includegraphics[width=\linewidth]{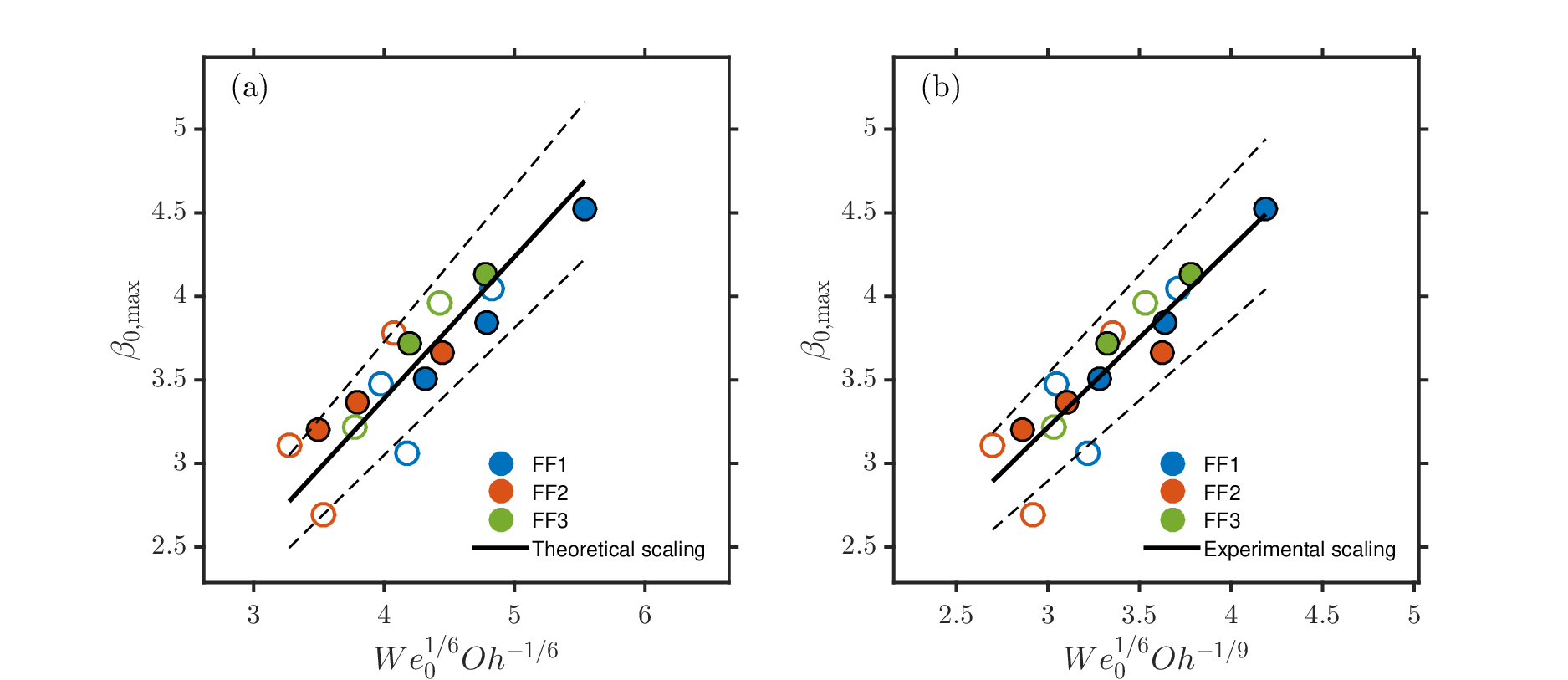}
    \caption{\protect\justifying Scaling of the zero-field maximum spreading factor
    $\beta_{0,max}$ against (a) the theoretical group
    $We_0^{1/6}Oh^{-1/6}$ of Liu \textit{et al.}~\cite{liu2025maximum}. The equation of the black solid line is $0.85 We_0^{1/6}Oh^{-1/6}$ with $R^2 = 0.76$ and (b) the freely fitted empirical group
    $We_0^{1/6}Oh^{-1/9}$. Here the equation of the black solid line is $1.07 We_0^{1/6}Oh^{-1/9}$ with $R^2 = 0.85$. Filled and open symbols denote the
    glass and PET substrates, respectively; colours denote FF1, FF2, and FF3. Dashed lines indicate $\pm10\%$ deviation from the solid fit line.}
    \label{fig:zerofield_spreading_scaling}
\end{figure}

The three ferrofluids used in the present study span exactly this low-to-moderate viscosity window with a range of $Oh$ falls under the above prescribed limit under no magnetic field. This motivates testing whether the
same $(We_0/Oh)^{1/6}$ scaling holds once the impact target is itself a liquid drop rather than a solid. Using the merged velocity
and diameter scales established in
Section~\ref{subsec:evolution of crown} ($V'_i$ and $D_t$), we define
the zero-field maximum spreading factor as $\beta_{0,max} =
D_{max}/D_i$ and test it against both the theoretical exponent of Liu \textit{et al.} and a free empirical fit to our own data. Fig.~\ref{fig:zerofield_spreading_scaling}a shows $\beta_{0,max}$
the plotted values $We_0^{1/6}Oh^{-1/6}$ for all three ferrofluids on both substrates. The data collapse onto the theoretical scaling to
within $\pm10\%$ across the full range tested.
Fig.~\ref{fig:zerofield_spreading_scaling}b repeats the comparison with a freely fitted power law, $\beta_{0,max} \sim
We_0^{1/6}Oh^{-1/9}$, obtained directly from our data. Surprisingly, the exponent on $We_0$ recovered from the fit agrees with both the theoretical value of $1/6$ and the independent empirical value of $0.166$ reported by Scheller and Bousfield for single-drop impact. But the exponent recovered for $Oh$, $-1/9$, is somewhat weaker than the theoretical $-1/6$. A likely reason is that coalescence itself consumes part of the impact energy before spreading even begins which has not been considered in the study of from Liu \textit{et al.}'s single-drop derivation. With less energy entering the spreading phase, the viscous dissipation that sets the $Oh$ dependence is correspondingly reduced, weakening the exponent relative to the single-drop prediction.
 
This agreement matters beyond simply confirming that the ferrofluids behave as Newtonian liquids in the absence of a field, which was already anticipated. It also establishes that the merged velocity and diameter scales, $V'_i$ and $D_t$, introduced to account for the momentum shared between the falling and sessile drops at coalescence are the physically correct scales to carry forward. This zero-field baseline is the reference against which the field-modified spreading and in particular, the saturation of $D_{max}$ at high field strengths and release heights already noted in Section~\ref{subsec:effect of magnetic field on maximum spreading} are assessed below.
\subsubsection{Scaling of maximum spreading diameter with magnetic field}
\label{subsubsec:Scaling of maximum spreading diameter with magnetic field}
The zero-field scaling derived above rests on the assumption, following Liu \textit{et al.}~\cite{liu2025maximum}, that viscous dissipation
is confined to a thin oscillatory boundary layer near the substrate. This assumption is not expected to survive once a field is applied. Ferrofluids are well known to develop a pronounced magnetoviscous
response under an applied field, as field-aligned particle chains resist the local shear associated with flow~\cite{odenbach2000magnetoviscous}, and we have shown comparable field-induced thickening for the present fluids in extensional flow~\cite{bera2026non}. With the effective viscosity of the spreading lamella itself elevated by the field, confining dissipation to a thin near-wall layer is no longer justified. Here, we follow Chandra and Avedisian~\cite{chandra1991collision}, who modelled viscous dissipation as occurring across the full height of the spreading lamella rather than within a boundary layer, and adapt
their approach to the present drop-on-drop, field-induced geometry.
 
The energy balance is constructed between the state immediately after coalescence and the state at maximum spreading. Now including
a magnetostatic energy term alongside the kinetic and surface energy terms used previously, the scale of initial energies are,
\begin{align}
    E_{ki} &\sim \rho V_i'^2 D_t^3, \\
    E_{si} &\sim \sigma D_t^2, \\
    E_{mi} &\sim -\frac{\chi B^2}{2\mu_0(1+N_i\chi)}\, \forall,
\end{align}
the kinetic, surface, and magnetostatic energies just after coalescence, where $\forall$ is the volume of the coalesced drop and $N_i$ is its demagnetization factor, appropriate to its near-spherical shape at this stage. At maximum spreading, the kinetic energy vanishes, and the scales surface and magnetostatic energies
become
\begin{align}
    E_{kf} &= 0, \\
    E_{sf} &\sim \sigma D_{max}^2, \\
    E_{mf} &\sim -\frac{\chi B^2}{2\mu_0(1+N_f\chi)}\, \forall,
\end{align}
with $N_f$ the demagnetization factor of the drop in its final, flattened pancake shape. Following Chandra and Avedisian~\cite{chandra1991collision}, the viscous dissipation is
estimated as
\begin{equation}
    E_{diss} \sim \mu\left(\frac{V_i'}{h_l}\right)^2 D_{max}^2 h_l
    \left(\frac{D_{max}}{2V_i'}\right),
\end{equation}
where $h_l$ (as shown in Fig.~\ref{fig:experimental_setup_magnetic_field}b) is the lamella height at maximum spreading and $D_{max}/(2V_i')$ is the characteristic spreading timescale. Conservation of volume between the coalesced drop and the spread
lamella,
\begin{equation}
    \frac{\pi}{6}D_t^3 \sim \frac{\pi}{4}D_{max}^2 h_l,
\end{equation}
gives $h_l \sim D_t^3/D_{max}^2$, so that the dissipation term reduces to,
\begin{equation}
    E_{diss} \sim \frac{\mu V_i' D_{max}^5}{D_t^3}.
\end{equation}
 
Equating the initial and final energy states, $E_{ki}+E_{si}+(E_{mi}-E_{mf})
\sim E_{sf}+E_{diss}$, and simplifying the magnetostatic term using
$N_f \ll N_i$ (the flattened final state has a negligible demagnetization factor compared with the near-spherical initial state), gives
\begin{equation}
    \rho V_i'^2 D_t^3 + \sigma D_t^2 + \frac{\chi B^2}{\mu_0}D_t^3
    \sim \sigma D_{max}^2 + \frac{\mu V_i' D_{max}^5}{D_t^3}.
\end{equation}
Dividing both sides by $\sigma D_t^2$ and writing $D_{max}=\beta_{max}D_i$, with $D_i \sim 2^{-1/3}D_t$ from volume conservation of the two
equal volume drops at coalescence, the ratio $D_i/D_t$ is an $O(1)$
constant that can be absorbed into the scaling. This gives
\begin{equation}
    We_m + 1 + Bo_m \sim \beta_{max}^2 +
    \frac{\mu V_i'}{\sigma}\beta_{max}^5,
\end{equation}
where $We_m = \rho V_i'^2 D_t/\sigma$ and $Bo_m = \chi B^2
D_t/(\mu_0\sigma)$ are the magnetic Weber and Bond numbers introduced in Section~\ref{subsec:evolution of crown}. In the asymptotic regime, where the inertial and magnetic driving terms dominate the initial surface energy, and where the viscous dissipation term dominates the final
surface energy term (i.e. $We_m + Bo_m \gg 1$ and the dissipation term
dominates $\beta_{max}^2$), this reduces to a balance between the  combined inertial plus magnetic energy and the viscous dissipation,
\begin{equation}
    We_m + Bo_m \sim \frac{\mu V_i'}{\sigma}\beta_{max}^5.
\end{equation}
Rewriting the right-hand side in terms of $Oh$ and $We_m$ we obtain, $We_m + Bo_m \sim Oh\sqrt{We_m}\,\beta_{max}^5$. So the theoretical scaling for the maximum spreading factor under a magnetic field is,
\begin{equation}
    \beta_{max} \sim \left(\frac{We_m + Bo_m}{Oh\sqrt{We_m}}\right)^{1/5}
    \equiv Z^{1/5}.
    \label{eq:betamax_field_theoretical}
\end{equation}
Equivalently as, $Oh\sqrt{We_m} = \mu V_i'/\sigma \equiv Ca_m$, the magnetic Capillary number built on the merged impact velocity $Vi'$, the above parameter can aslo be written as,  
\begin{equation}
    Z = \frac{We_m+Bo_m}{Ca_m}.
\end{equation}
Here this is the ratio of combined intertial and magnetic driving forces to the viscous resistance opposing the spreading.
\begin{figure}
    \centering
    \includegraphics[width=\linewidth]{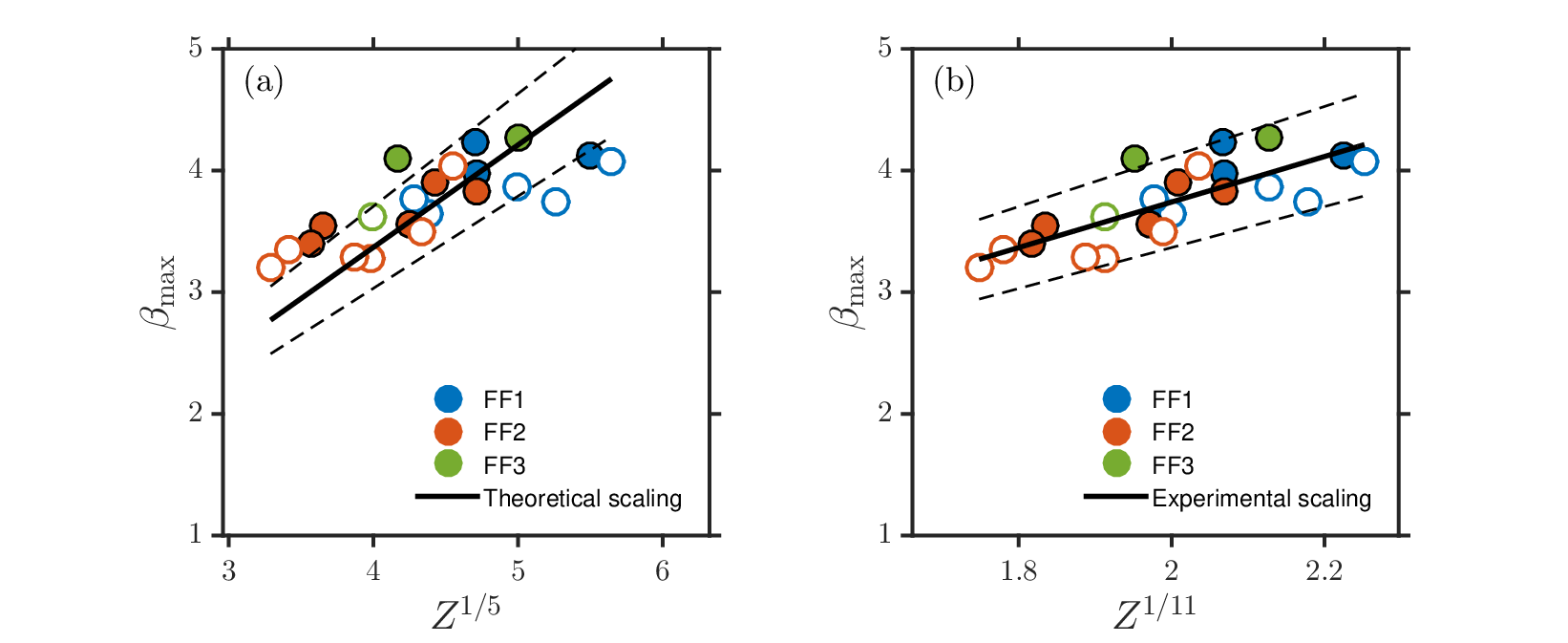}
    \caption{\protect\justifying (a)Maximum spreading factor $\beta_{max}$ under a magnetic
    field plotted against the theoretical group $Z^{1/5}$. The equation of the black solid line is $0.84Z^{1/5}$ with $R^2 = 0.71$ and (b) $\beta_{max}$ under a magnetic
    field plotted against the same group $Z$, with a freely fitted
    exponent. Here the equation of the black solid line is $1.87Z^{1/11}$ with $R^2 = 0.85$. Filled and open symbols denote the
    glass and PET substrates, respectively; colours denote FF1, FF2, and FF3. Dashed lines indicate $\pm10\%$ deviation from the solid fit line.}
    \label{fig:maxspreading_withfield}
\end{figure}

Fig.~\ref{fig:maxspreading_withfield}a tests this scaling with field maximum spreading paremeter, and the collapse is noticeably poorer than that obtained for the zero-field case. 
Retaining the same composite group $Z$ but fitting the exponent freely to the data, rather than fixing it
at the theoretical value of $1/5$, produces a substantially better collapse, shown in Fig.~\ref{fig:maxspreading_withfield}b. The
fitted exponent, however, is only $\approx 1/11$, little less than half the theoretical value. Since the derivation leading to
Eq.~\ref{eq:betamax_field_theoretical} balances the driving energy against viscous dissipation, an exponent weaker than predicted indicates that the actual dissipation is larger than the model
accounts for at a given value of $Z$. The model, built on the zero-field Ohnesorge number $Oh$, is systematically underestimating how dissipative the spreading lamella becomes once a field is present.

This deficit can be quantified directly by returning to the full energy balance and obtaining the value of the Ohnesorge number it would require, given the measured $We_m$, $Bo_m$, and $\beta_{max}$ for each condition. By retaining the surface energy terms, we obtain,
\begin{equation}
    We_m + 1 + Bo_m \sim \beta_{max}^2 + Oh_{eff}\sqrt{We_m}\,\beta_{max}^5.
\end{equation}
$Oh_{eff}$ reflects the extra viscous
dissipation caused by the field, which was not considered in the earlier approximation. Solving for
$Oh_{eff}$ gives,
\begin{equation}
    Oh_{eff} \equiv \frac{We_m + 1 + Bo_m - \beta_{max}^2}{\sqrt{We_m}\,\beta_{max}^5},
    \label{eq:Oheff}
\end{equation}
which isolates the additional dissipation associated specifically
with the field.
 
\begin{figure}
    \centering
    \includegraphics[width=0.6\linewidth]{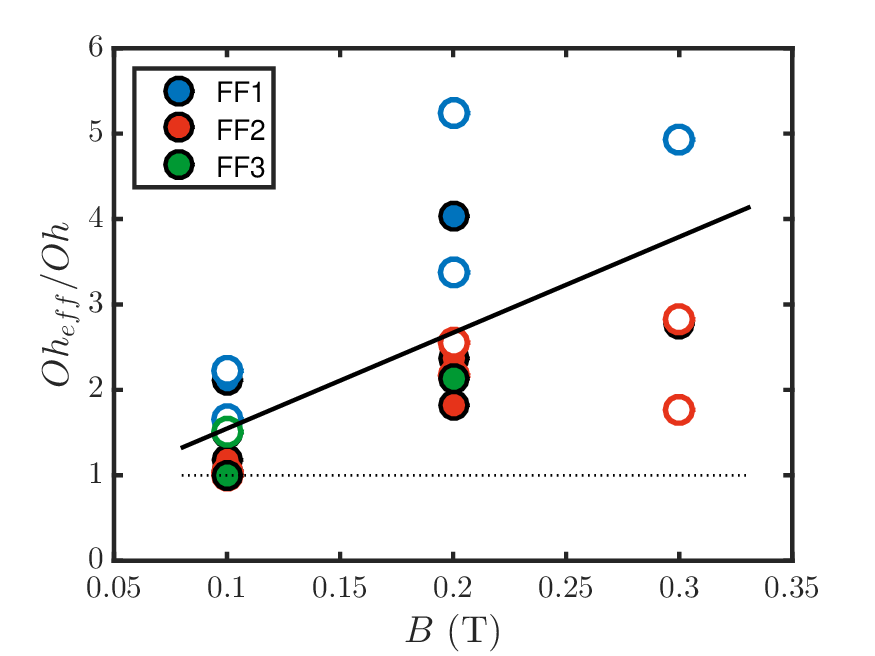}
    \caption{\protect\justifying Effective Ohnesorge number, normalized by its zero-field value, plotted against magnetic field strength $B$ for all three
    ferrofluids and both substrates. The black solid line is a guide to the eye indicating the increase in $Oh_{eff}$ with increasing magnetic field.}
    \label{fig:Oheff_vs_field}
\end{figure}
 
Fig.~\ref{fig:Oheff_vs_field} shows $Oh_{eff}$, computed from Eq.~\ref{eq:Oheff} using the measured spreading data and normalized by its zero-field value, plotted against the applied field strength.
Across all three ferrofluids and both substrates. $Oh_{eff}/Oh$ rises well above unity as the field strength increases, in some cases
exceeding the zero-field value by a factor of four to five at
$0.2$--$0.3$~T. This is the direct signature of the magnetoviscous effect~\cite{rosensweig1969viscosity,pop2005microstructure}. The field increases the effective viscosity of the spreading
lamella, and does so more strongly as the field is increased, exactly the mechanism missing from Eq.~\ref{eq:betamax_field_theoretical}, which treats $Oh$ as fixed at its zero-field value. Because the theoretical scaling underestimates
dissipation more severely at higher field, where $Oh_{eff}$ departs furthest from $Oh$.
 
This also completes the explanation, deferred from
Section~\ref{subsec:effect of magnetic field on maximum spreading},
for why $D_{max}$ was observed to saturate rather than continuing to increase at higher field strengths. The field
assists spreading through two channels captured explicitly in
Eq.~\ref{eq:betamax_field_theoretical}. It raises the impact velocity $V_i'$ via the Kelvin force (Section~\ref{subsec:effect of magnetic field on impact velocity}), and it contributes directly to the driving energy through the magnetic Bond number $Bo_m$. At the same time, the field is making the spreading lamella itself more viscous, as Fig.~\ref{fig:Oheff_vs_field} shows, and that extra viscosity works against the spreading the field is driving. This answers the reason behind the plateau in $D_{max}$ seen in Fig.~\ref{fig:spreading_dynamics}b. This is
where these two effects catch up with each other. The field adds energy, but the added resistance cancels it out.
\subsubsection{Rim detachment phenomena and it's criterion}
\label{subsubsec:Rim detachment phenomena and it's criterion}
At zero field, for every tested release height, the crown that forms after coalescence always retracts and collapses back onto the spreading lamella. Once a field of at least $0.1$~T is applied, however, a qualitatively different outcome appears at sufficiently large release heights (from $32$~cm onward, for the fluids and
substrates studied): rather than retracting, the rim of the crown detaches entirely from the sheet beneath it. This is not simply a stronger version of ordinary crown retraction; it is a distinct rim
instability. A stronger field raises the impact velocity $V_i'$ through the Kelvin force
(Section~\ref{subsec:effect of magnetic field on impact velocity}), and a larger impact velocity produces a taller crown (Section~\ref{subsec:evolution of crown}). Detachment occurs once the inertia carried by the rim, set by this impact velocity, grows large enough to overcome the surface tension holding the rim to the sheet below i.e., once the kinetic energy of the rim exceeds the surface energy resisting its separation. As the field is increased
further beyond this threshold, the detached rim does not detach only, it fragments into a train of daughter droplets, a second, independent instability triggered once the rim is no longer anchored to the retracting sheet.
 
\begin{figure*}[]
    \centering
    \begin{minipage}{0.48\textwidth}
        \centering
        \includegraphics[width=\textwidth]{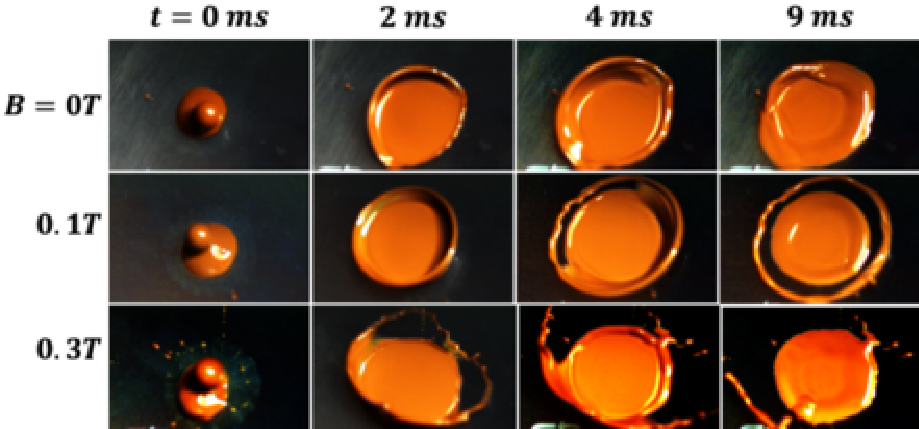}
        \vspace{0.15em}
        \textbf{(a)}
    \end{minipage}
    \hfill
    \begin{minipage}{0.48\textwidth}
        \centering
        \includegraphics[width=\textwidth]{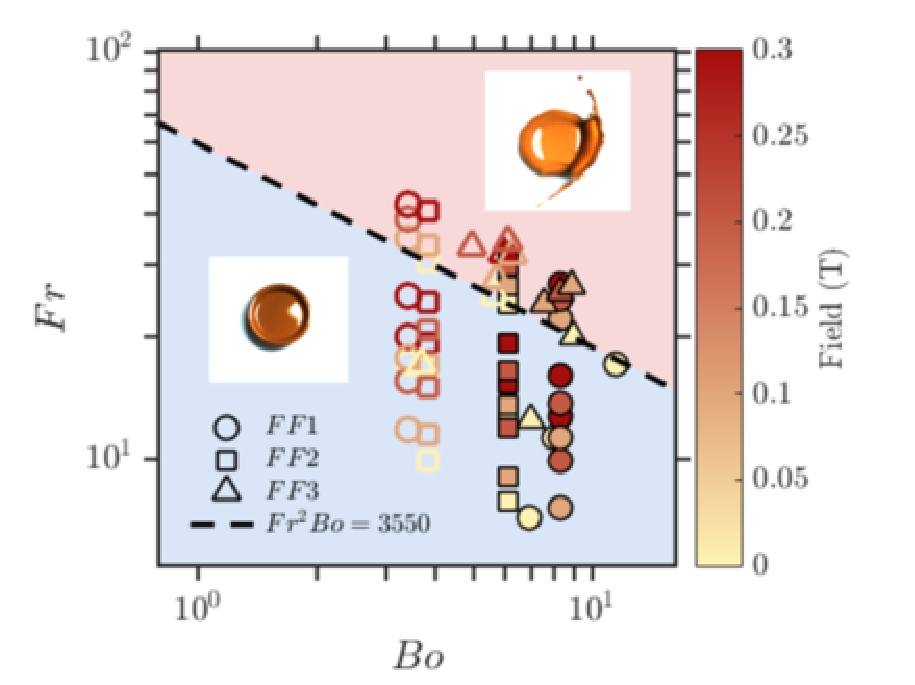}
        \vspace{0.15em}
        \textbf{(b)}
    \end{minipage}
    \caption{\protect\justifying (a) Snapshots of a falling FF1 drop impacting a sessile FF1 drop from $32 cm$ height on a PET substrate at $t=0$, $2$, $4$, and $9$~ms, for $B=0$, $0.1$, and $0.3$~T. The rim remains attached at
    $B=0$~T, detaches cleanly at $0.1$~T, and fragments into daughter droplets at $0.3$~T. (b) Regime map for rim detachment in the $Fr$--$Bo$ plane. The dashed line, $Fr^2 Bo = 3550$, separates conditions where the rim remains attached to the retracting sheet (below the line) from those where it detaches (above the line).
    Circles, squares and triangles  denote FF1, FF2 and FF3, respectively, coloured by the applied field strength. Filled and open symbols are used to denote glass and PET substrate respectively. }
    \label{fig:rim_detachment}
\end{figure*}
 
To make this competition between inertia and surface tension quantitative, we work with the Froude number, $Fr = V_i'^2/(gD_t)$,
already used to scale the crown height, and the (gravitational) Bond number, $Bo = \rho g D_t^2/\sigma$, which compares the weight of the
coalesced drop to the surface tension holding it together. Note that
this $Bo$ is distinct from the magnetic Bond number $Bo_m$ used
earlier in Section~\ref{subsubsec:Scaling of maximum spreading diameter with magnetic field}; here it enters only through its usual gravitational definition. For each fluid, substrate, and field strength, we identify the critical Froude number, $Fr_c$, at which rim detachment first appears at a given $Bo$. Plotting $Fr_c$ against $Bo$ for all
conditions tested shows that these critical points collapse onto a single curve when combined as $Fr_c^2 Bo$, which is found empirically
to equal $3550$ across the dataset, with approximately $11\%$ scatter. The detachment criterion is therefore
\begin{equation}
    Fr^2 Bo \sim \frac{\rho V_i'^4}{g\sigma} = 3550.
    \label{eq:rim_detachment_criterion}
\end{equation}
The right-hand form follows directly from substituting the definitions of $Fr$ and $Bo$. The length scale $D_t$ cancels exactly
between the two numbers, so that $Fr^2 Bo$ depends only on the impact
velocity and the fluid properties $\rho$, $g$, and $\sigma$, and not on the size of the coalesced drop. This is itself a useful result, the detachment threshold is set entirely by how the impact velocity
competes against surface tension, independent of $D_t$ and it is precisely the ratio of an inertial energy scale, $\rho V_i'^4/g$, to
a capillary energy scale, $\sigma$, that decides whether the rim separates. It is worth being clear the fields enter the the above criterion indirectly. Here the magnetic field helps to set $V_i'$, through Eq.~\ref{eq:impactveloexpre}, where height and field are added together. So rim detachment is not really a magnetic effect at its core, it is a velocity threshold and field is one way to reach it. A drop falling from a sufficient height with no field should cross the same threshold on gravity alone. We don't observe such a phenomena at zero field even at our maximum height because the upper height was bounded for practical reasons (see Section~\ref{sec:method}).
 
Fig.~\ref{fig:rim_detachment}a shows the snapshots of the falling FF1 drop on the PET substrate from $32 cm$ height for a representative case. At $B=0$~T the rim
stays attached throughout the event. At $B=0.1$~T, just past the detachment threshold, the rim separates cleanly from the sheet. At $B=0.3$~T, well above threshold, the same detachment is accompanied by visible fragmentation of the rim into a train of smaller daughter droplets. Fig.~\ref{fig:rim_detachment}b shows this regime map directly, with the dashed line $Fr^2Bo = 3550$ separating the two outcomes. Below
the line, the rim remains attached and retracts with the collapsing crown and above it, the rim detaches.
 
This finding also closes a question left open in
Section~\ref{subsec:effect of magnetic field on maximum spreading}:
why $D_{max}$ was observed to saturate with increasing release height at fixed field, rather than continuing to grow, in
Fig.~\ref{fig:spreading_dynamics}c. Once with applied field, the release height is large enough to cross the rim detachment threshold of
Eq.~\ref{eq:rim_detachment_criterion}, a fraction of the drop's mass, is ejected away as detached rim and, at even higher fields, as fragmented daughter droplets, rather than contributing to further radial spreading. Thus, with increasing height, the maximum spread diameter $D_{max}$ ceases to increase in the presence of magnetic field, which provides a second, independent mechanism for the saturation of $D_{max}$,
distinct from the earlier magnetoviscous resistance case.
\section{Conclusion}
\label{sec:conclusion}
This work looked at what happens when a falling ferrofluid drop lands on a sessile drop of the same liquid under a horizontal magnetic field, a case that sits between two problems usually
studied apart: drop-on-drop impact, and magnetically controlled ferrofluid spreading. We followed the impact in the order it
actually unfolds, crown formation, maximum spreading, and rim detachment  across three ferrofluids, two substrates, and a range
of heights and field strengths.

For the crown, a simple ballistic launch sets an upper bound, $H_{c,max}/D_t \sim Fr$. The real crown falls well below this bound, as it should once surface energy, viscosity, and rim mass
accumulation, all left out of the ballistic picture on purpose, are brought back in. We find an empirical scaling as, $Fr^{0.5\pm0.02}$ for maximum crown height. For spreading, the zero-field data collapse onto the boundary-layer
scaling $\beta_{0,max}\sim(We_0/Oh)^{1/6}$ already known for single-drop impact, which tells us the merged scales $V_i'$ and $D_t$ are the right variables to carry forward. Once a field is
switched on, a bulk-dissipation energy balance predicts $\beta_{max}\sim Z^{1/5}$, but the data fit closer to $Z^{1/11}$, weaker than predicted because the field itself makes the fluid more viscous, which we capture through an effective Ohnesorge number $Oh_{eff}$ that climbs to four or five times its zero-field value at
high field. The field drives spreading and resists it at the same time, causing it to impede the  increment of $D_{max}$ further with increasing magnetic field. Finally, the rim detaches only past a threshold field and height, following one clean criterion, $Fr^2Bo\approx3550$, independent of drop size. This is the point where the crown's inertia beats the surface tension holding its rim in place. Beyond this threshold the detached rim breaks into
daughter droplets, carrying away mass and momentum that would otherwise have kept the lamella spreading.

These results matter wherever a field is used to steer repeated drop impacts, such as magnetic spray coating, 3D printing, and microfluidic or biomedical droplet handling cases that often land drops on liquid left by earlier drops, or need a field as the only handle once a drop has left the nozzle. The scalings here: crown height, field-dependent spreading, and the 
$Fr^2Bo$ criterion, help to predict whether such an impact spreads cleanly, throws up a crown, or sheds satellite droplets, wherever print quality or coating uniformity depends on avoiding stray splashing. Throughout we have had to infer the internal physics, such as viscous dissipation, the growth of field-induced particle chains, the flow field during spreading indirectly, through scaling arguments, and quantities like $Oh_{eff}$, rather than observing it directly. As a scope of future work, complete numerical simulations of field-induced drop-on-drop phenomena will enable us to resolve the internal velocity field and spatial distribution of dissipation that the present experiments could only scale within physically consistent bounds. 

\begin{acknowledgments}
NSB and AKJ thanks IIT Kharagpur for the PhD scholarship. PD thanks the Anusandhan National Research Foundation (ANRF) for funding this research (project: ANRF/ARG/2025/000031/ENS).
\end{acknowledgments}

\nocite{*}

\bibliography{apssamp}

\end{document}